\documentclass[12pt,psfig,showpacs,letterpaper]{revtex4}
\usepackage{geometry}
\usepackage{graphicx}
\usepackage{natbib}
\usepackage{amsmath}
\usepackage{amssymb}
\begin{document}
\bibliographystyle{acm}
\title{SELF HEATING OF CORONA BY ELECTROSTATIC FIELDS DRIVEN BY SHEARED FLOWS}
\author{H. Saleem$^{1}$, S. Ali$^{2}$, S. Poedts$^{3}$\\1\&2. National Centre for Physics,
Quaid-i-Azam University Campus,\\ Shahdarra Valley Road, Islamabad,
Pakistan \\3.  K.U.Leuven, Belgium}
\begin{abstract}
A mechanism of self-heating of solar corona is pointed out. It is
shown that the free energy available in the form of sheared flows
gives rise to unstable electrostatic waves which accelerate the
particles and heat them. The electrostatic perturbations take place
through two processes (a) by purely growing sheared flow-driven
instability and (b) by sheared flow-driven drift waves. These
processes occur throughout the corona and hence the self-heating is
very important in this plasma. These instabilities can give rise to
local electrostatic potentials $\varphi$ of the order of about 100
volts or less within $3\times10^{-2}$ to a few seconds time if the
initial perturbation is assumed to be about one percent that is
$\frac{e\varphi}{T_{e}}\simeq10^{-2}$. The components of wave
lengths in the direction perpendicular to external magnetic field
$\textbf{B}_{0}$ vary from about 10m to 1m. The purely growing
instability creates electrostatic fields by sheared flows even if
the density gradient does not exist whereas the density gradient is
crucial for the concurrence of drift wave instability.\\
\textit{Subject headings:} Sun: self-heating of corona, sheared
flow-driven instability, drift waves.
\end{abstract}
\maketitle

\large{

\section{Introduction}Several theoretical models have been
presented to explain the cause of solar coronal heating (Aschwanden
2001; Mandrini, Demoulin, \& Klimchuk 2000). It is a well-known
counter institutive fact that the corona is 200 times hotter than
the chromosphere while it is a rarefied collision-less plasma with
electron temperature $T_{e}\simeq10^{6}(K)$. There are many puzzles
in the problem of coronal heating. The temperature rises by two
orders of magnitude from upper chromosphere to lower corona through
a transition layer of only about 500 km (Priest 1982; Narain;
Ulmsch- neider \& Klimchuk 2006). The proposed wave heating
mechanisms assume that the waves originate in the lower regions and
deposit their energy in the corona. Observational data of Skylab,
Yohkoh, SOHO and TRACE has indicated that the entire corona is
filled with open and closed magnetic field lines and only a subset
of it is loaded with hot plasma at a given time (Litwin \& Rosner
1993; Hara et al. 1992; Moses et al. 1997;
Schrijver et al. 1999). The coronal loops have also been considered to be responsible for localized heating.\\
The bright coronal loops have higher densities compared to the
ambient faint coronal plasma which indicates that the heated plasma
originates from the dense
chromosphere. The coronal heating by acoustic waves originating from
global oscillations have been ruled out (Aschwanden 2001). Alfven
waves have been considered to be the best candidate for carrying
adequate energy fluxes from chromosphere to corona (Hollweg \&
Sterling 1994). A great deal of work on Alfven wave heating of the
corona has appeared in literature (Ionson 1983; Mok 1987;
Steinolfson \& Davila 1993; Ofman, Davila \& Steinolfson 1995;
Helberstadt \& Goedbloed
1995; Ruderman et al. 1997).\\
Any unstable wave in this plasma can cause coronal heating through
damping either by wave-particle interaction or by other linear and
nonlinear mechanisms. Most of the theories pertaining to wave
heating in corona are based on magnetohydrodynamics (MHD) and hence
electrostatic drift-waves have not been investigated while in fact
the system itself is highly inhomogeneous. There are many suggested
mechanisms for heating in which the main energy source is provided
to corona by other regions. \\We point out an important energy
source available within the corona to excite short scale
electrostatic perturbations and that are the sheared flows of plasma
streams and rivers. These electrostatic perturbations can be
associated with either the purely unstable mode (D'Angelo 1965) or
the drift-waves of two fluid plasma depending upon the scales of
density inhomogeneity and the wavelengths. Vranjes and Poedts
(2009a) have proposed a new paradigm of solar coronal heating by
drift waves. They have used the results of kinetic theory which
predicts that drift wave is a universally unstable mode and density
gradient is the source of its instability which is also the cause of
its existence. It has been proposed that the drift waves can heat
the corona through two possible ways, one due to the Landau damping
effects in the direction parallel to the magnetic field, and another
one, the stochastic heating in the perpendicular direction. It has
been confirmed (De Pontieu et al. 2007) that the solar atmosphere is
highly structured and has inhomogeneous density filaments of various
sizes. Therefore drift waves in corona may have several different
scales of wavelengths.\\Vranjes and Poedts (2009a) assume external
magnetic field is along z-axis while the density gradient is along
x-axis and wave propagates under local approximation in yz-plane
with perpendicular wavelength $\lambda_{y}=\frac{2\pi}{k_{y}}$ and
parallel wavelength $\lambda_{z}=\frac{2\pi}{k_{z}}$ while
$k_{z}<<k_{y}$ (where $k_{y}$ and $k_{z}$ are respective wave
numbers). The drift waves have been considered with
$\lambda_{y}\simeq0.5m$ while $\lambda_{z}=20km$ and $L_{n}=100m$
(where $L_{n}=\left|\frac{1}{n_{0}}\frac{dn_{0}}{dx}\right|^{-1}km$)
is the density gradient scale length. \\We focus our attention on
the fact that sheared flows are omnipresent in the corona. Thus a
large source of energy for the excitation of short scale
electrostatic perturbations exist within the corona. Two types of
electrostatic instabilities are expected to
occur throughout the corona continually:\\
I. Purely growing sheared flow-driven instability in doppler shifted
frame (D'Angelo 1965) which exist even if the plasma density is
uniform\\
II. Drift wave instability which needs plasma density gradient for
it's existence (Kadomtsev \& Timofeev (1963)) and sheared flows for instability (Saleem, Vranjes \& Poedts (2007))\\
We investigate the drift waves having relatively longer wavelengths
compared to ion Larmor radius  so that the fluid model can be
justified. Moreover, the electrons are assumed to follow Boltsmann
distribution. Therefore frequencies  $\omega_r$ of drift waves must
fulfill the condition $\omega_r << \nu_{te} k_z$ where $\nu_{te} =
(T_e /m_e)^{1/2}$ is the electron thermal velocity and $k_z$ is the
wave number parallel to external magnetic field. The drift waves and
purely growing D'Angelo mode have shorter wavelengths compared to
Alfven waves of MHD model, in general. The electrostatic
perturbations driven by sheared flows can transfer their energy to
plasma particles. For the case of drift waves a detailed picture has
been presented by Vranjes \& Peodts (2009a). The drift dissipative
instabilities have also been discussed by Vranjes \& Peodts (2009b).

\section{Theoretical Model} Let us consider the
collisionless coronal plasma consisting of two types of ions j = a,
b, where 'a' represents Hydrogen ions and 'b' represents Helium
ions. We choose the ambient magnetic field as
$\textbf{B}_{0}=B_{0}\mathbf{\hat{z}}\equiv$constant, $\nabla
n_{j0}=-\mathbf{\hat{x}}\left|\frac{dn_{j0}}{dx}\right|$, and the
shear flow as
$\nu_{0}(x)=v_{j0}(x)\mathbf{\hat{z}}=v_{0}(x)\mathbf{\hat{z}}$
where $v_{0}$ is the same for all species; electrons and both type
of ions. Each species has a zero-order diamagnetic drift
$\textbf{v}_{Dj0}=-\left(\frac{T_{j}}{q_{j}B_{0}}\right)\nabla ln
n_{j0}\times\mathbf{\hat{z}}$ where $T_{j}$ is temperature and
$q_{j}$ is charge of the jth species of ions. For electrostatic
perturbation, the momentum equation yields the perpendicular
component of the velocity of jth species of ions as,
$$\textbf{v}_{\bot
j}=\frac{1}{B_{0}}(\textbf{E}_{\bot}\times\mathbf{\hat{z}})-\frac{T_{j}}{q_{j}B_{0}}(\nabla
ln
n_{j}\times\mathbf{\hat{z}})-\frac{1}{q_{j}B_{0}}\left(\frac{\nabla.\mathbf{\Pi}_{j}}{n_{j}}\right)-\frac{1}{\Omega_{j}}(\partial_{t}
+\textbf{v}_{j}.\nabla)\textbf{v}_{j}\times\mathbf{\hat{z}}$$
$$=\textbf{v}_{E}+\textbf{v}_{Dj}+\textbf{v}_{\Pi
j}+\textbf{v}_{pj}\eqno(1)$$ where $\textbf{v}_{E}, \textbf{v}_{Dj},
\textbf{v}_{\Pi j}$ and $\textbf{v}_{pj}$ are the electric,
diamagnetic, stress tensor and polarization drifts, respectively.
The parallel component of momentum equation gives,
$$(\partial_t + v_{j0z}
\partial_z)v_{jz1} + v_{jz1}
d_{x} \nu_{joz} (x) = \frac{q_j}{m_j} E_{z1} - \frac{T_{j0}}{m_j
n_{j0}}
\partial_z n_{j1} \eqno{(2)}$$ Here $d_{x}=\frac{d}{dx}$ and the subscripts
naught (0) and one (1) denote zero order and linear quantities,
respectively. The continuity equation for jth ions is,
$$(\partial_t n_{j1} + v_{0jz})
\partial_t n_{j1} + \nabla n_{j0} . \textbf{v}_{E} +
\frac{n_{j0}}{B_0 \Omega_j} (\partial_t + v_{j0z}
\partial_z)\nabla_{\perp} . \textbf{E}_{\perp}$$ $$\nu_{0z}-\frac{T_j}{q_j
B_0 \Omega_j} (\partial_t + \nu_{j0z} \partial_z) \nabla_{\perp}^{2}
n_{j1} + n_{j0} \partial_z v_{jz1} = 0 \eqno{(3)}$$  where $\Omega_j
= \frac{q_j B_0}{m_j}$ and $\nabla \nu_{0z} (x) = + \hat{\mathbf{x}}
\left|\frac{d\nu_0 (x)}{dx}\right|$. In obtaining (3), we have used
the following relation
$$\nabla.[n_{j}.(\textbf{v}_{jp}+\textbf{v}_{j\pi})]=\frac{n_{j0}}{B_{0}\Omega_{j}}\partial_{t}(\nabla_{\bot}.\textbf{E}_{\bot1})-\frac{T_{j}}{q_{j}B_{0}\Omega_{j}}\partial_{t}\nabla^{2}n_{j1}$$
$$+\frac{n_{j0}}{\Omega_{j}}v_{0}(x)\partial_{z}\left\{\frac{1}{B_{0}}\nabla_{\bot}.\textbf{E}_{\bot}-\frac{T_{j}}{q_{j}B_{0}}\frac{\nabla^{2}_{\bot}n_{j1}}{n_{j0}}\right\}\eqno(4)$$
The continuity equations for j = a, b can also be expressed as,
$$R^{2}_{a}n_{a1}=-n_{a0}S^{2}_{a}\Phi_{1}\eqno(5)$$
$$R^{2}_{b}n_{b1}=-n_{b0}S^{2}_{b}\Phi_{1}\eqno(6)$$ where
$$R^{2}_{a}=(1+\rho^{2}_{aT}k^{2}_{y})\Omega^{2}_{\omega}-\nu^{2}_{te}k^{2}_{y}$$
$$R^{2}_{b}=(1+\rho^{2}_{bT}k^{2}_{y})\Omega^{2}_{\omega}-\nu^{2}_{te}k^{2}_{y}$$
$$S^{2}_{a}=\left\{-\omega^{*}_{a}\Omega_{\omega}+\rho^{2}_{as}k^{2}_{y}\Omega^{2}_{\omega}+c^{2}_{as}k_{y}k_{z}\frac{1}{\Omega_{a}}\frac{dv_{0}}{dx}-c^{2}_{as}k^{2}_{z}\right\},$$
$$S^{2}_{b}=\left\{-\omega^{*}_{b}\Omega_{\omega}+\rho^{2}_{bs}k^{2}_{y}\Omega^{2}_{\omega}+c^{2}_{bs}k_{y}k_{z}\frac{1}{\Omega_{b}}\frac{dv_{0}}{dx}-c^{2}_{bs}k^{2}_{z}\right\},$$
Here we have defined $\nu^{2}_{Tj}=\frac{T_{j0}}{m_{j}}$,
$\Omega_{j}=\frac{q_{j}B_{0}}{m_{j}}$,
$c^{2}_{js}=\frac{T_{e}}{m_{j}}$,
$\Omega_{\omega}=(\omega-\omega_{0})$, $\omega_{0}=v_{0}k_{z}$,
$\omega^{*}_{j}=D_{e}\kappa_{jn}\kappa_{y}$,
$\kappa_{jn}=\left|\frac{1}{n_{j0}}\frac{dn_{j0}}{dx}\right|$, and
$\Phi=\frac{e\varphi}{T_{c}}$. \\Poisson equation in this case can
be written as,
$$\nabla.\textbf{E}_{1}=\frac{e}{\epsilon_{0}}(n_{a1}+n_{b1}-n_{e1})\eqno(7)$$
and electrons (e) are assumed to follow the Boltzmann relation,
$$n_{e1}\simeq n_{e0}e^{\Phi}\eqno(8)$$ In steady state the relation
$n_{e0}=n_{a0}+n_{b0}$ holds. Equations (5 - 8) yields a fourth
order dispersion relation as follows,
$$L_{4}\Omega^{4}_{\omega}+L_{3}\Omega^{3}_{\omega}+L_{2}\Omega^{2}_{\omega}+L_{2}\Omega^{2}_{\omega}+L_{1}\Omega_{\omega}+L_{0}=0\eqno(9)$$
where
$$L_{4}=\Lambda^{2}\alpha^{2}_{a}\alpha^{2}_{b}+\frac{n_{a0}}{n_{e0}}(\alpha^{2}_{b}\rho^{2}_{as}k^{2}_{y})+\frac{n_{b0}}{n_{e0}}(\alpha^{2}_{a}\rho^{2}_{bs}k^{2}_{y})$$
$$L_{3}=-\left\{\frac{n_{a0}}{n_{e0}}\alpha^{2}_{b}\omega^{*}_{a}+\frac{n_{b0}}{n_{e0}}\alpha^{2}_{a}\omega^{*}_{b}\right\}$$
$$L_{2}=-\Lambda^{2}(\alpha^{2}_{a}\nu^{2}_{bT}k^{2}_{z}+\alpha^{2}_{b}\nu^{2}_{aT}k^{2}_{z})+\frac{n_{a0}}{n_{e0}}(\alpha^{2}_{b}g_{a}c^{2}_{sa}k^{2}_{z}$$
$$-\rho^{2}_{as}k^{2}_{y}\nu^{2}_{bT}k^{2}_{z})+\frac{n_{b0}}{n_{e0}}(\alpha^{2}_{a}g_{b}c^{2}_{bs}k^{2}_{z}-\rho^{2}_{bs}k^{2}_{y}\nu^{2}_{aT}k^{2}_{z})$$
$$L_{1}=\frac{n_{a0}}{n_{e0}}(\nu^{2}_{bT}k^{2}_{z}\omega^{*}_{a})+\frac{n_{b0}}{n_{e0}}(\nu^{2}_{aT}k^{2}_{z}\omega^{*}_{b})$$
$$L_{0}=\Lambda^{2}\nu^{2}_{aT}k^{2}_{z}\nu^{2}_{bT}k^{2}_{z}-\frac{n_{a0}}{n_{e0}}g_{a}\nu^{2}_{bT}k^{2}_{z}c^{2}_{as}k^{2}_{z}-\frac{n_{b0}}{n_{e0}}g_{b}\nu^{2}_{aT}k^{2}_{z}c^{2}_{bs}k^{2}_{z}$$
Here $g_{j}=\left(\frac{k_{y}}{k_{z}}A_{j}-1\right),
A_{j}=\frac{1}{\Omega_{j}}\frac{dv_{0}}{dx},
\alpha^{2}_{j}=(1+\rho^{2}_{jT}k^{2}_{y}),
\lambda^{2}_{De}=\frac{\epsilon_{0}T_{e}}{n_{e0}e^{2}}$, and
$\Lambda^{2}=(1+\lambda^{2}_{De}k^{2})$.
\section{Application to Corona}
Now we show that the theoretical model presented above is perfectly
applicable to coronal plasma. Let us choose the parameters of the
corona (Priest 1982) as $n_{e0}=10{15}m^{-3}$, $n_{a0}=0.9n_{e0}$,
and $T_{e}=10^{6}K$. Since the condition $T_{e}<T_{H}<T_{He}$ holds
in this plasma (Hansteen, Leer, and Holtzer 1997), therefore we
assume $T_{a}=2.5 T_{e}$ and $T_{b}=3T_{e}$. \\First we show that
the sheared flows of electron proton plasma of corona gives rise to
purely growing and oscillatory drift wave instabilities. Then it
will be shown that the presence of second ion (Helium 10\%) in this
plasma modifies the growth rate and the real frequencies of the
drift wave in different parameter regimes. If this 10\%
concentration of the ions in the plasma is neglected, then we may
use $n_{b0}=0$ and $T_{b0}=0$ in the equation (9). In the limit
$\lambda^{2}_{De}k^{2}<<1$, we have $\Lambda^{2}=1$,
$\alpha^{2}_{b}=1$, $\frac{n_{a0}}{n_{e0}}=1$, $\nu^{2}_{bT}=0$ and
it gives $L_{4}=\rho^{2}_{as}k^{2}_{y}+\alpha^{2}_{a}$,
$L_{3}=-\omega^{*}_{a}$,
$L_{2}=-nu^{2}_{aT}k^{2}_{z}+g_{a}c^{2}_{as}k^{2}_{z}$ and
$L_{1}=0=L_{0}$. In this case equation (9) reduces to
$$(1+\rho^{2}_{aT}k^{2}_{y}+\rho^{2}_{as}k^{2}_{y})\Omega^{2}_{\omega}-\omega^{*}_{a}\Omega_{\omega}+A_{a}k_{z}c^{2}_{as}
-(c^{2}_{as}+v^{2}_{aT})k^{2}_{z}=0\eqno(10)$$ This quadratic
equation has two roots
$$(\Omega_{\omega})_{1,2}=\frac{1}{2\Lambda_{0}}\left[\omega^{*}_{a}\pm\left\{(\omega^{*}_{a})^{2}
+4\Lambda_{0}c^{2}_{as}k^{2}_{z}\left((1+\sigma_{a})-A_{a}\frac{k_{y}}{k_{z}}\right)\right\}\right]$$
$$(\Omega_{\omega})_{1,2}=\frac{1}{2\Lambda_{0}}\left\{\omega^{*}_{a}\pm\left[(\omega^{*}_{a})^{2}
+4\Lambda_{0}c^{2}_{as}k^{2}_{z}\left((1+\sigma_{a})-A_{a}\frac{k_{y}}{k_{z}}\right)\right]\right\}\eqno(11)$$
where
$\Lambda_{0}=(1+\rho^{2}_{aT}k^{2}_{y}+\rho^{2}_{as}k^{2}_{y})$.
Following conditions should satisfy simultaneously for the
instability. $$(1+\sigma_{a})\frac{k_{z}}{k_{y}}<A_{a}\eqno(12)$$
and
$$(\omega^{*}_{a})^{2}<4\Lambda_{0}c^{2}_{as}k^{2}_{z}\left|\left((1+\sigma_{a})-A_{a}\frac{k_{y}}{k_{z}}\right)\right|\eqno(13)$$
We shall assume that $\kappa_{nj}$ has the same value for all
species and that can be denoted by $\kappa_{n}$. In Fig. 1, the
growth rates of the shear flow-driven electrostatic instability are
plotted using equation (10) against different gradient scale lengths
of the plasma flow parallel to external magnetic field in case of
homogeneous density. The real frequency in laboratory frame is
$\omega_{0}$. But in the moving frame of plasma, it is a purely
growing instability and therefore $\omega_{i}$ has been plotted vs
perpendicular wavenumber $k_{y}$. We notice that the imaginary
frequency $\omega_{i}$ decreases corresponding to the same $k_{y}$
when the shear flow gradient scale length
$L_{v}=\frac{1}{\kappa_{v}}$ increases. Thus the steeper gradients
give rise to larger growth rate of the instability. \\In the
presence of density gradient, the electrostatic drift wave can also
be excited. Fig. 2 shows that the drift wave becomes unstable for
$2<k_{y}<2.5$ for a very short range of wavelengths. Somewhere in
between $k_{y}=2$ and $k_{y}=2.5$, the real frequency of drift wave
$\omega^{*}_{a}$ is larger than the growth rate $\omega_{i}$ that is
$\omega_{i}<\omega^{*}_{a}$ and hence linear analysis is valid. The
drift wave seems to be driven by the sheared-flow. The pattern of
growth rate $\omega_{i}$ for $2<k_{y}<2.5$ becomes different in Fig.
2 which has $\kappa_{n}\neq0$ compared to Fig. 2 plotted for
$\kappa_{n}=0$. For $2.5\leq k_{y}$, the purely growing instability
dominates because here $\omega_{r}<<\omega_{i}$ and hence it shows
the shear flow-driven instability with local real frequency
$\omega_{0}=\nu_{0}k_{z}$. Both $\omega_{i}$ and $\omega_{r}$
becomes larger for drift wave when the shear flow gradient is
steeper as shown in Fig. 3 while the density gradient is kept
constant $\kappa_{n}=1.9\times10^{-3}m^{-1}$. We notice that the
growth rates and real frequency of the drift wave decrease
corresponding to smaller $k_{y}$ and $k_{z}$ compared to Fig. 2
which is quite natural. In this case the instabilities require
larger value of flow because $c^{2}_{as}k^{2}_{z}$ - term becomes
smaller in condition (13). In Fig. 4, the effect of the ration of
$\frac{k_{z}}{k_{y}}$ on instabilities is shown. In Fig. 5 the
effect of the presence of helium ions on real and imaginary
frequencies is shown on the shear flow-driven instability (for
$\kappa_{n}=0$) and on the drift wave instability for
$(\kappa_{n}\neq0)$.

\section{Role of Dissipation} The Solar corona is commonly assumed
to be a collision-less plasma because most of the wave studies deal
with the relatively higher frequency Alfven waves using MHD
equations or ion acoustic waves with the wave numbers larger than
our regime of parameters and hence the frequency becomes larger than
the electron-ion collision frequency $\nu_{ei}$. The drift waves
have been investigated in corona first time (to the best of authors
knowledge) by Vranjes and Poedts (2009a) and (2009b). The drift
waves investigated by Vranjes \& Poedts (2009a) have frequencies
$\omega_{r}\sim\omega_{i}\sim10^{2}rad/s$ and
$\nu_{ei}\simeq30rad/s$ (Vranjes \& Poedts 2009b). Therefore the
ideal plasma approximation is valid. In the higher frequency
($\omega{*}_{a}$) regime it does not seem preferable to assume
electrons to be inertia-less because the condition
$\omega^{*}_{a}<<v_{te}k_{z}$ may not remain valid. We choose the
wave parameters such that the condition $\omega^{*}_{a},
\omega^{*}_{b}<<v_{te}k_{z}$ remains valid and electrons follow the
Boltzmann distribution. But here we see another small effect namely
the dissipation. Let us look at the Fig. 2(a) to analyze the role of
drift wave in corona. This wave is stable for $k_{y}<2$ and then for
$2<k_{y}<3$, the wave develops $\omega_{i}$ and for $3<k_{y}$ we can
see that $\omega_{r}<\omega_{i}$ and the shear flow-driven
instability dominates. To understand the wave behaviour, we choose a
value of $k_{y}$ in between 2 and 3. Let $k_{y}=2.2 m^{-1}$, then
for $\kappa_{n}=1.9\times10^{-3}m^{-1}$ we find
$\omega^{*}_{a}=D_{e}\kappa_{n}k_{y}\simeq36 rad/s$ which is equal
to $\nu_{ei}$. Therefore, dissipation can play some role in this
frequency regime. \\It is important to note that the well-known
drift dissipative instability (DDI) (Weiland 2000) does not become
important even in this range of frequencies because
$\omega^{*}_{a}\nless \nu_{ei}$. The parallel momentum equation for
electrons yields (Weiland 2000)
$$\frac{n_{e1}}{n_{e0}}\simeq\frac{e\varphi}{T_{e}}\left\{1-\iota\frac{\nu_{ei}}{\nu^{2}_{te}k^{2}_{z}}(\omega{*}_{a}-\omega)\right\}\eqno(14)$$
in the limit $\omega^{*}_{a}<<\nu_{ei}$. In ion continuity equation
the limit $c^{2}_{s}k^{2}_{z}<<\omega^{*}_{a}$ is used along with
$n_{e}\simeq n_{i}$, to obtain linear dispersion relation for DI
with the real frequency $\omega_{rd}$ and imaginary frequency
$\omega_{id}$ as (Weiland 2000),
$$\omega_{rd}\simeq\frac{\omega^{*}_{a}}{(1+\rho^{2}_{as}k^{2}_{y})}\eqno(15)$$
and
$$\omega_{id}\simeq\left(\frac{\omega^{2}_{r}}{\nu^{2}_{te}k^{2}_{z}}\right)\nu_{ei}\rho^{2}_{as}k^{2}_{y}\eqno(16)$$
In our case, the drift wave frequency can be of the order of
$\nu_{ei}$. Since we have shown that the drift wave becomes unstable
due to sheared flow in the absence of electron-ion collisions,
therefore in our parameter regime, the dissipation can just add its
small effect to the already unstable perturbation. But the DDI is
not applicable because the real frequency $\omega^{*}_{a}$ is not
much smaller than $\nu_{ei}$.
\\The Fig. 2(a) indicates that for small $k_{y}-$values, the drift
wave is stable and shear flow does not have an effect on it. So for
longer wavelengths $k_{y}<<2$, the drift dissipative instability in
corona can take place. As an example, let us choose
$k_{y}=0.1m^{-1}$ and take rest of the parameters to be the same
i.e. $\kappa_{n}=1.9\times10^{-3}$, $k_{z}=10^{-4}k_{y}$,
$B_{0}=10^{-2}T$ etc. Then we find $\omega^{*}_{a}\simeq1.63$,
$\nu_{te}k_{z}\simeq36$ while $\nu_{ei}\simeq36$. Then
$\omega^{*}_{a}\simeq\nu_{ei}<<\Omega_{i}$ holds along with
$\omega^{*}_{a}<<\frac{\nu^{2}_{te}k^{2}_{z}}{\nu_{ei}}$. Therefore,
the drift dissipative instability in corona gives rise to drift
waves having very low frequency $\omega^{*}_{a}\simeq1Hz$ and
relatively longer wavelength
$\lambda_{y}\simeq\frac{2\pi}{k_{y}}\simeq60m$.
\section{Discussion} It has been proposed that the large perturbed
electrostatic fields are generated throughout the solar corona, due
to localized sheared plasma flows, which accelerate the particles
and heat them. The corona is not a static ball of plasma, rather it
has flows and gradients. Therefore, it has already been proposed
that the free energy available in the form of density gradients can
produce electrostatic drift waves (Vranjes \& Poedts 2009a, 2009b).
Using the results of kinetic theory, these authors have shown that
the universally unstable drift waves can heat the coronal ions very
efficiently. The density gradient in the direction perpendicular to
the external magnetic field is the cause of the existence of these
waves as well as it is the source for their instability. The waves
transfer their energy to plasma particles through Landau damping
(the wave-particle interaction). This process cannot be studied
using fluid models. \\But the two fluid theory has predicted a
Kelvin-Helmholtz type instability which takes place in plasmas
because of sheared flows (D'Angelo 1965). It shows that if both
electrons and ions flow with the same velocity along the external
magnetic field $\textbf{B}_{0}$ and there exists a gradient in flow
in the direction perpendicular to $\textbf{B}_{0}$, then the
perturbed electrostatic field becomes unstable. These are purely
growing fields in the frame of reference of the flow wide spectrum
of wavelengths. But in laboratory frame these unstable perturbations
have associated local real frequencies as
$\omega_{0}=v_{0}(x)k_{z}$. The sheared flows in corona can give
rise to two types of electrostatic instabilities continuously
throughout the corona.
\begin{enumerate}
\item Sheared flow-driven instability (D'Angelo 1965) even if the
plasma density is uniform $(\kappa_{n}=0)$
\item The drift wave, which exists if the density is non-uniform $\kappa_{n}\neq0$,
instability due to sheared flow
\end{enumerate}
These instabilities have been investigated neglecting the effects of
electron-ion collisions. For $k_{y}\simeq10m^{-1}$, we find that the
sheared flow-driven instability can create potential
$\varphi\simeq68$ volt in about growth-time $\tau_{g}\simeq0.03s$ if
at $t=0$ we assume $\frac{e\varphi}{T_{e}}\simeq10^{-2}$. Note that
Vranjes \& Poedts (2009a) have estimated that large frequency drift
waves ($\omega_{r}\simeq 2.5\times 10^{2}$) can give rise to this
value of potential in 0.02 s. The components of wavelength chosen in
the perpendicular and parallel directions are, respectively,
$\lambda_{y}=0.5m^{-1}$ and $\lambda_{z}=20km$. Then they have
$\nu_{te}k_{z}=1.2\times10^{3}$. \\We have also shown that the drift
waves having $\omega_{r}<50$ rad/s satisfy
$\omega_{r}<<\nu_{te}k_{z}$ can become unstable due to sheared flow.
The collisional effects have been neglected which can only modify
the growth rates by small amounts. It is well-known that the
electron-ion collisions can drive drift dissipative instability and
it does not require plasma flow to occur. The collisions produce
electrostatic drift waves having very small real frequency
$\omega_{rd}\simeq1.6$ rad/sec corresponding to
$k_{y}\simeq0.1m^{-1}$ and longer wavelengths
$\lambda_{y}=\frac{2\pi}{k_{y}}\simeq60m$. If the initial
perturbation is assumed to be
$\Phi_{0}=e^{\frac{e\varphi}{T_{e}}}=10^{-2}$, then $\varphi$ will
take time $\tau_{g}\simeq15$ minutes to grow up to $\varphi=86$
volt. But smaller values of $\varphi$ will be produced in much small
times than 15 minutes. Thus we conclude that drift waves of
different frequencies and wavelengths are produced in the solar
corona due to sheared flows and electron-ion collisions. \\Thus
electrostatic fields are almost omnipresent in the coronal plasma
and continuous self-heating is taking place due to sheared flows and
density gradients. The present investigation shows that even if the
plasma density is uniform in a region, the electrostatic fields will
be produced because corona is not static and sheared flows occur
everywhere. Therefore, the sheared flow driven instability plays an
important role in self-heating of the corona.
\begin{figure}
\center
\includegraphics[scale=0.7]{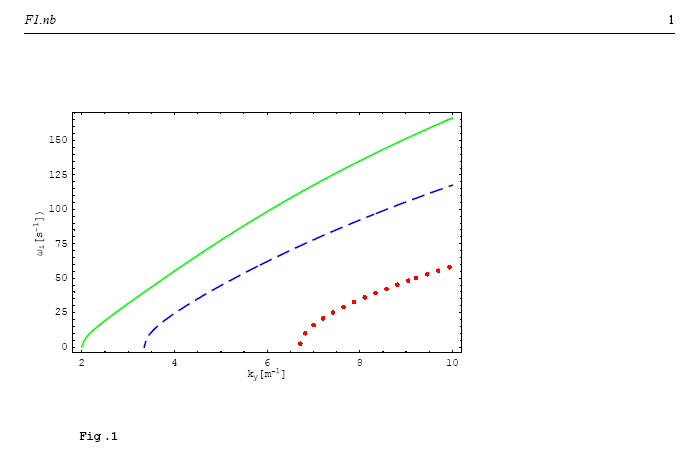}
\caption{(Color online) The growth rate $(\omega_{i})$ is plotted
against the perpendicular component of the wavenumber $(k_{y})$ for
different inverse
velocity scalelengths; $\kappa_{v}=k_{y}/60$ (solid curve), $\kappa_{v}%
=k_{y}/100$ (dashed curve), and $\kappa_{v}=k_{y}/200$ (dotted
curve) with $n_{e0}\sim n_{a0}\sim10^{15}$ \textrm{m}$^{-3}$,
$T_{e}=10^{6}$ \textrm{K, }$T_{a}=2.5$ $T_{e}$, $B_{0}$\
$\sim10^{-2}$ \textrm{Tesla}, $v_{0}=10$
\textrm{k}$\mathrm{m/s,}$ $\kappa_{nj}=0,$ $n_{b0}=0,$ and $k_{z}=10^{-4}%
k_{y}$ $\mathrm{m}^{-1}.$}
\end{figure}

\begin{figure}
\center
\includegraphics[scale=0.7]{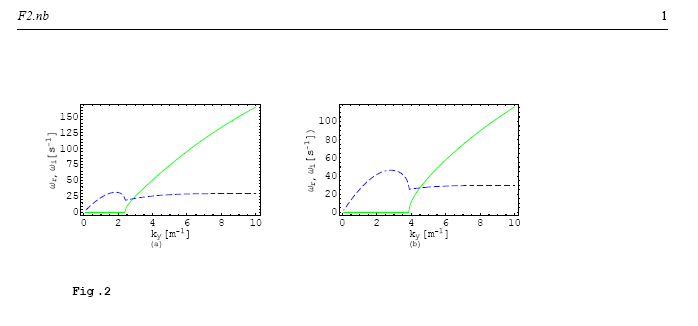}
\caption{(Color online) The real and imaginary frequencies
[$\omega_{r}$ $($dashed curve$),$ $\omega_{i}$ $($solid curve$)$]
are plotted against the perpendicular component of the wavenumber
$(k_{y})$ for varying inverse
velocity scalelength (a) $\kappa_{v}=k_{y}/60,$ and (b) $\kappa_{v}%
=k_{y}/100,$ taking $v_{0}=10$ \textrm{k}$\mathrm{m/s,}$ $\kappa_{n}%
=1.9\times10^{-3}$ $\mathrm{m}^{-1},$ $n_{b0}=0,$ and $k_{z}=10^{-4}%
k_{y}\mathrm{m}^{-1}.$ All other parameters are the same as in
Fig.1.}
\end{figure}

\begin{figure}
\center
\includegraphics[scale=0.7]{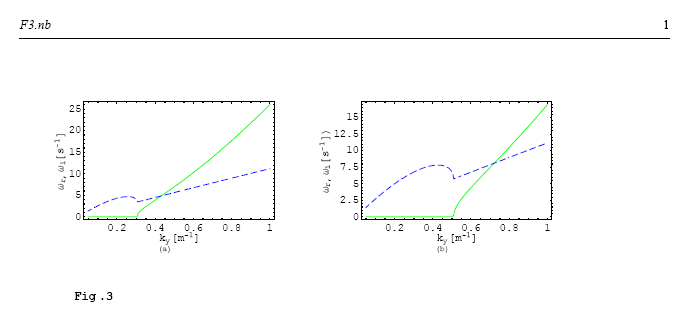}
\caption{(Color online) The real and imaginary frequencies
[$\omega_{r}$ $($dashed curve$),$ $\omega_{i}$ $($solid curve$)$]
are plotted against the perpendicular component of the wavenumber
$(k_{y})$ for changing inverse velocity scalelength (a)
$\kappa_{v}=k_{y}/60$ and (b) $\kappa_{v}=k_{y}/100,$ keeping
$v_{0}=70$ \textrm{k}$\mathrm{m/s,}$ $\kappa_{n}=1.9\times10^{-3}$
$\mathrm{m}^{-1},$ $n_{b0}=0,$ and
$k_{z}=10^{-4}k_{y}\mathrm{m}^{-1}.$ All other parameters are the
same as in Fig.1.}
\end{figure}

\begin{figure}
\center
\includegraphics[scale=0.7]{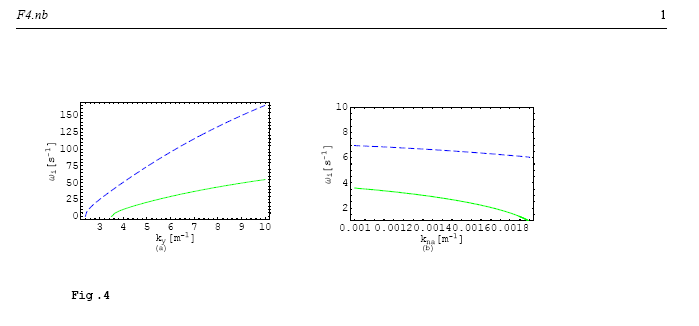}
\caption{(Color online) The growth rate $\omega_{i}$ is plotted
against the perpendicular component of the wavenumber $(k_{y})$ and
the inverse density scalelength ($\kappa_{n})$, respectively,
varying the parallel component of the wavenumver (a)
$k_{z}=10^{-5}k_{y}\mathrm{m}^{-1}$\textrm{ }(solid curve),\textrm{
}$k_{z}=10^{-4}k_{y}\mathrm{m}^{-1}$ (dashed curve) with fixed
values of $v_{0}=10$ \textrm{k}$\mathrm{m/s,}$
$\kappa_{n}=1.9\times10^{-3}$ $\mathrm{m}^{-1},$ and
$\kappa_{v}=k_{y}/60,$ and the streaming velocity (b)
$v_{0}=50$ \textrm{k}$\mathrm{m/s}$ (solid curve), $v_{0}=70$ \textrm{k}%
$\mathrm{m/s}$ (dashed curve), with $\kappa_{v}=k_{y}/60,$
$k_{y}=0.5$ $\mathrm{m}^{-1}$, and
$k_{z}=10^{-4}k_{y}\mathrm{m}^{-1}.$ All other parameters are the
same as in Fig.1.}
\end{figure}

\begin{figure}
\center
\includegraphics[scale=0.7]{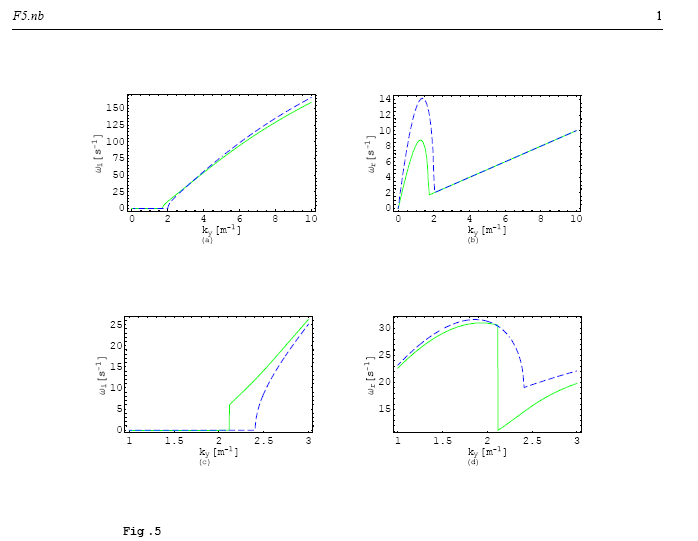}
\caption{(Color online) The imaginary and real frequencies (a)
$\omega_{i}$ and (b) $\omega_{r},$ are plotted against the
perpendicular component of the wavenumber $(k_{y}),$ respectively,
for $n_{b0}=0$ $($dashed curve$)$ and
$n_{b0}=0.1n_{e0}$ $($solid curve$),$ with $v_{0}=10$ \textrm{k}%
$\mathrm{m/s,}$ $n_{e0}\sim10^{15}$ \textrm{m}$^{-3}$,
$n_{a0}\sim0.9n_{e0}$ and $B_{0}$\ $\sim10^{-2}$ \textrm{Tesla},
taking $\kappa_{n}=1.9\times
10^{-3}\mathrm{m}^{-1},$ $\kappa_{v}=k_{y}/60,$ and $k_{z}=10^{-4}%
k_{y}\mathrm{m}^{-1}$.}
\end{figure}

\pagebreak \textbf{Captions}\\

Figure 6: (Color online) The imaginary and real frequencies (a)
$\omega_{i}$ and (b) $\omega_{r},$ are plotted against the
perpendicular component of the wavenumber $(k_{y}),$ respectively,
for $n_{b0}=0$ $($dashed curve$)$ and
$n_{b0}=0.1n_{e0}$ $($solid curve$),$ with $v_{0}=10$ \textrm{k}%
$\mathrm{m/s,}$ $n_{e0}\sim10^{15}$ \textrm{m}$^{-3}$,
$n_{a0}\sim0.9n_{e0}$ and $B_{0}$\ $\sim10^{-2}$ \textrm{Tesla},
taking $\kappa_{n}=0,$ $\kappa _{v}=k_{y}/60,$ and
$k_{z}=10^{-4}k_{y}\mathrm{m}^{-1}$.

\pagebreak

}
\end{document}